\documentclass[preprint,showpacs,preprintnumbers,amsmath,amssymb]{revtex4}

\usepackage{graphicx}

\begin{document}

\title{The Antiferromagnetic Heisenberg Model on Clusters with Icosahedral
       Symmetry}

\author{N.P. Konstantinidis$^{*}$}
\affiliation{Department of Physics and Department of Mathematics, University of
             Dublin, Trinity College, Dublin 2, Ireland}

\date{\today}

\begin{abstract}
The antiferromagnetic Heisenberg model is considered for spins
$s_{i}=\frac{1}{2}$ located on the vertices of the dodecahedron and the
icosahedron, which belong to the point symmetry group $I_{h}$. Taking into
account the permutational and spin inversion symmetries of the Hamiltonian
results in a drastic reduction of the dimensionality of the problem, leading to
full diagonalization for both clusters. There is a strong signature of the
frustration present in the systems in the low energy spectrum, where the first
excited states are singlets. Frustration also results in a doubly-peaked
specific heat as a function of temperature for the dodecahedron. Furthermore,
there is a discontinuity in the magnetization as a function of magnetic field
for the dodecahedron, where a specific total spin sector never becomes the
ground state in a field. This discontinuity is accompanied by a
magnetization plateau. The calculation is also extended for $s_{i}=1$ where
both systems again have singlet excitations. The magnetization of the
dodecahedron has now two discontinuities in an external field and also
magnetization plateaux, and the
specific heat of the icosahedron a two-peak structure as a function of
temperature. The similarities between the two systems suggest that the
antiferromagnetic Heisenberg model on a larger cluster with the same symmetry,
the $60$-site cluster, will have similar properties.
\end{abstract}


\maketitle

\section{Introduction}
\label{sec:1}
The antiferromagnetic Heisenberg model has been the object of intense
investigation in recent years as a prototype of strongly correlated electronic
behavior \cite{Manousakis91}. The effects of low dimensionality, quantum
fluctuations and frustration combine together to produce new phases different
from conventional Ne\'el-like order, where the order parameter is not the
staggered magnetization or there is a lack of a local order parameter
altogether \cite{Misguich03,Lhuillier02,Lhuillier01}. They can
also have dramatic consequences on the energy spectrum, as is the case for
frustrated multiple spin exchange models and the Kagom\'e lattice, where the
low energy excitations are singlets \cite{Waldtmann98}. Specific heat
calculations of small Kagom\'e lattice samples have revealed a two-peak
structure, with the first peak below the singlet-triplet gap
\cite{Sindzingre00,Syromyatnikov03}. The double-peak structure was also
obtained for other multi-spin exchange models \cite{Roger90,Misguich98}, a
pyrochlore slab \cite{Kawamura02}, and the $\Delta$ chain \cite{Kubo93}. It is
a natural question to ask whether there are other frustrated systems with such
unconventional behavior.

\begin{figure}
\includegraphics[width=5in,height=3.5in]{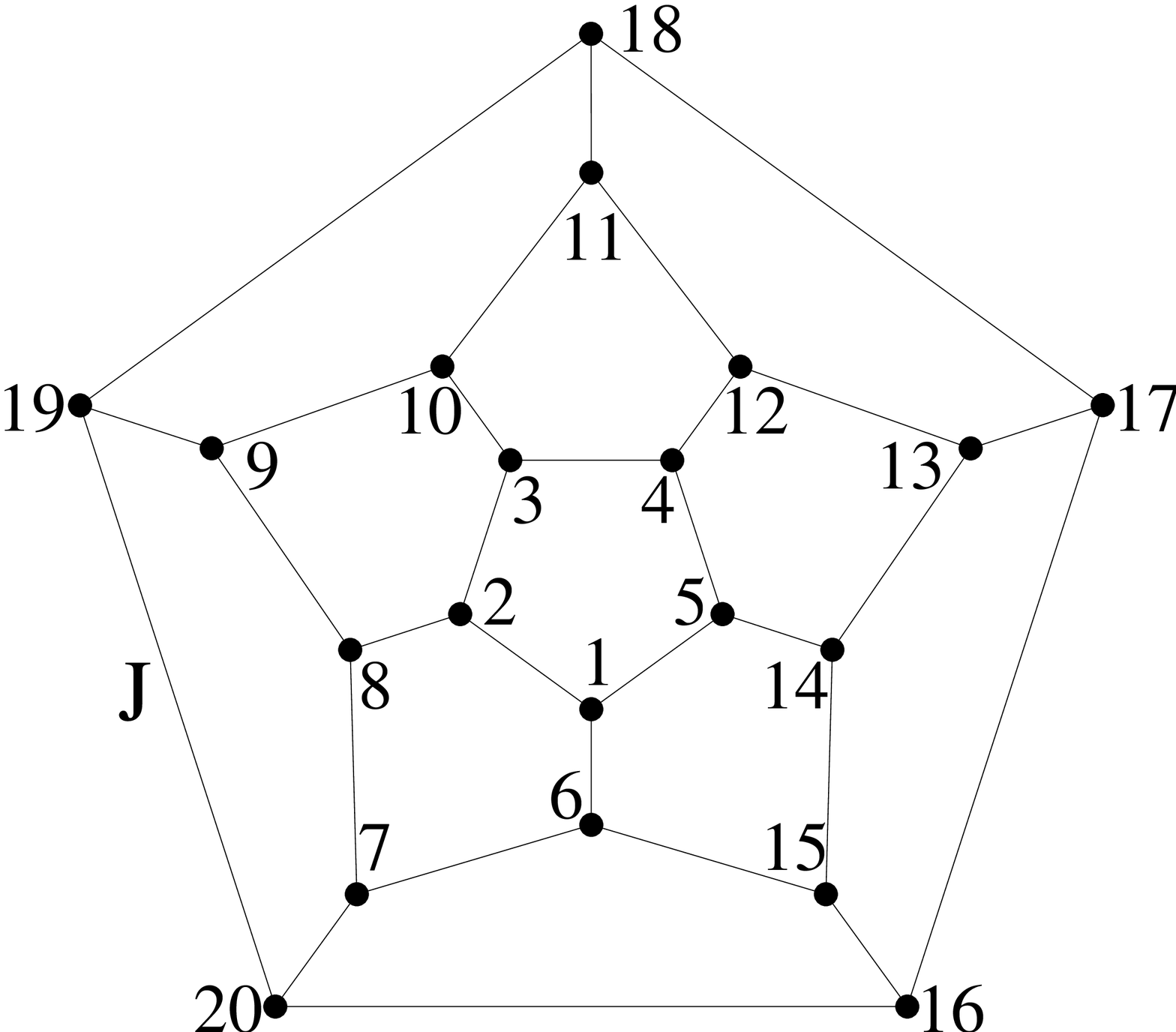}
\caption{Projection of the dodecahedron on a plane. The solid lines are
antiferromagnetic bonds $J$. The black circles are spins $s_{i}$.}
\label{fig:1}
\end{figure}

Here clusters with the connectivity of the fullerenes will be
considered \cite{Kroto85,Fowler95}. These molecules weakly
bound to form crystals which become superconductors when doped
with alkali metals \cite{Hebard91,Holczer91}. Their transition
temperature is above 40 K,
a transition temperature much higher than the one of conventional
superconductors. Chakravarty and Kivelson suggested that an electronic
mechanism at intermediate scales is responsible for superconductivity
in C$_{60}$, the fullerene with $60$ carbon atoms, when doped with alkali
metals \cite{Chakravarty01}. It is an open question if a
repulsive interaction can produce pairing in such systems.
The Hubbard model
has been used to investigate an electronic mechanism for superconductivity
on this cluster \cite{Ojeda99,Gonzalez96}. An exact treatment of the model
in the full Hilbert space of
the molecule is prohibitive due to its size. Therefore, as a first
step, smaller molecules of the fullerene type were considered to gain
insight, as well as the strong on-site repulsion limit of the Hubbard model at
half-filling, the antiferromagnetic Heisenberg model \cite{Coffey92,Trugman92}.
Coffey and Trugman found that connectivity and frustration lead to
non-trivial behavior at the classical level in a magnetic field
\cite{Coffey92}. To study the effect of quantum fluctuations on the classical
results a $20$-site cluster, the dodecahedron, was considered (figure
\ref{fig:1}). It is three-fold coordinated, has all sites equivalent and
consists of $12$ pentagons. The model was studied with perturbation
theory around the classical limit and was found to possess a singlet as the
first excited state, and a discontinuity in the magnetization as a function of
quantum fluctuations \cite{NPK01}. This discontinuity was originally found at
the classical level \cite{Coffey92}.
Therefore it is of interest to study the structure of the low energy spectrum
of the dodecahedron and its response in a magnetic field for
$s_{i}=\frac{1}{2}$. A more general question is if the presence of singlets in
the excitation spectrum is also a property of other clusters, and if there is a
correlation of magnetic behavior with space group symmetry and connectivity
\cite{NPK05-2}. The determination of the full energy spectrum will reveal if
the double-peak specific heat structure is also a property of the dodecahedron.
However,
the projection of the total spin on the $z$ axis, $S^{z}$, is not a good
quantum number when perturbing around the classical state. In addition,
multiprecision arithmetic was needed to analytically continue the series
expansions in \cite{NPK01}. Thus the full calculation of the energy spectrum
was
prohibited due to memory requirements, even though symmetry was partially taken
into account. The ground state properties of the model were firstly
calculated by Modine and Kaxiras \cite{Modine96}.

\begin{figure}
\includegraphics[width=5in,height=3.5in]{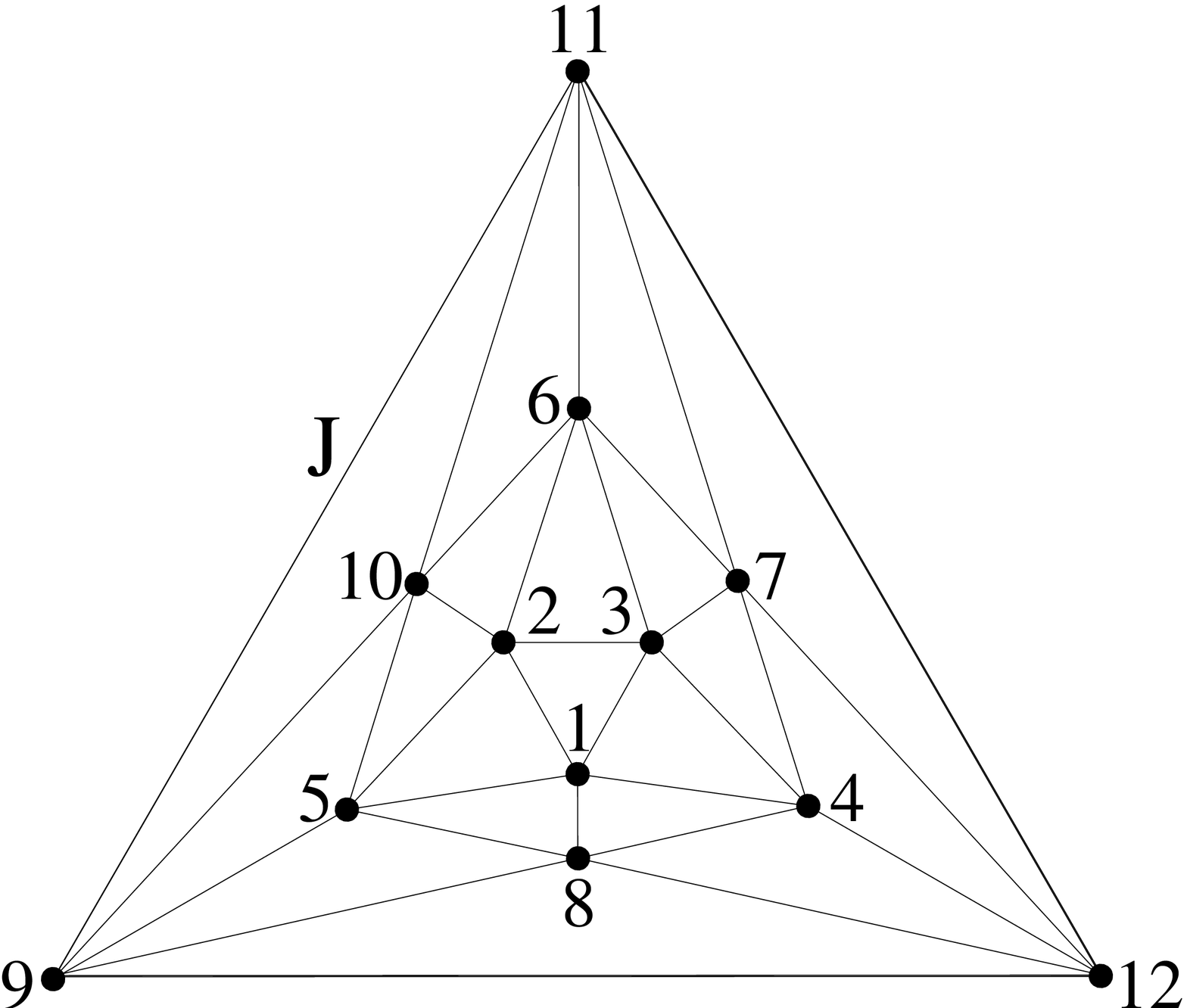}
\caption{Projection of the icosahedron on a plane. The solid lines are
antiferromagnetic bonds $J$. The black circles are spins $s_{i}$.}
\label{fig:2}
\end{figure}

In this paper the antiferromagnetic Heisenberg model is studied on the
dodecahedron and a smaller cluster with $12$ equivalent sites and the same
spatial symmetry, the icosahedron (figure \ref{fig:2}), for $s_{i}=\frac{1}{2}$
and $1$. Permutational and spin symmetries are taken into account
\cite{Bernu94,Florek02,Waldmann00}, and this leads to full diagonalization
except for the dodecahedron when $s_{i}=1$, where Lanczos diagonalization is
used \cite{ARPACK,Shaw}. The symmetry group of the clusters is $I_{h}$, the
largest point symmetry group with $120$ operations \cite{Altmann94}.
The icosahedron has $20$ triangles, and the spins on the vertices are five-fold
coordinated. Therefore the two clusters share the same symmetry but their
connectivity is different. The results for the low energy spectrum of the
dodecahedron for $s_{i}=\frac{1}{2}$ show that the ground state is a singlet,
while the first excited state is also a singlet but five-fold degenerate. The
next excited state is also a singlet, and then a series of states with
total spin $S=1$
follows. This series of low-lying singlets is a consequence of the connectivity
and frustration of the model. The specific heat also shows a
non-conventional behavior as a function of temperature, with a peak
inside the energy gap and a second peak at higher temperatures. The
low energy spectrum of the icosahedron is similar with the one of the
dodecahedron, with the same ordering and relative spacing of low energy levels.
The specific heat has a well-pronounced peak inside the energy gap,
but there is no second peak but rather a shoulder at higher temperatures.

For spins with magnitude $s_{i}=1$, the ground and first excited state of
the dodecahedron are closely spaced singlets. The next excited state is a
three-fold degenerate triplet. The low energy spectrum for the icosahedron is
again the same in the symmetry, the ordering and the relative spacing of the
low lying levels. The specific heat can only be calculated for the icosahedron
and has now two peaks, however there is no peak inside the energy
gap. This along with the reduced number of low energy singlet states
compared with the full quantum $s_{i}=\frac{1}{2}$ case, indicates a
change in the spectrum with increasing $s_{i}$.

The behavior of the magnetization in a magnetic field is non-trivial for the
dodecahedron. There is a discontinuity as the energy of a particular
total spin $S$ sector
never becomes the ground state in a field. This feature, observed
at the classical level by Coffey and Trugman \cite{Coffey92}, survives
in both the
$s_{i}=\frac{1}{2}$ and $1$ cases, twice in the latter. Similar behavior is not
observed for the
icosahedron, which also has discontinuous magnetization in a field at
the classical level. Even though the two clusters have the same spatial
symmetry, the behavior in a field appears to depend on their polygon
structure. The magnetization discontinuities in the dodecahedron are
accompanied by magnetization plateaux.

The similarities of the dodecahedron and icosahedron spectra suggest that
predictions
about the low energy structure can be made for a larger cluster of fullerene
type connectivity with the same symmetry $I_{h}$, the $60$-site cluster, where
again all sites are equivalent \cite{Fowler95}. Even though the $12$- and
$20$-site clusters have different coordination number and consist of different
types of polygons, the structure of the low energy spectrum is the same. The
$60$-site cluster has also a discontinuity in the magnetization in a
field at
the classical level \cite{Coffey92}. However, it is not clear what the response
in a magnetic field will be for the quantum case, since the two smaller
clusters have different behavior, which seems to depend on the connectivity as
well as their polygon structure. Similar considerations could correlate the
behavior of more complicated models with orbital degrees of freedom
like the Hubbard model on the
$60$-site cluster with the behavior of the same models on the dodecahedron and
the icosahedron.

The plan of this paper is as follows: in section \ref{sec:2} the model and
method are introduced, in section \ref{sec:3} the low energy spectra of the two
clusters are presented for both $s_{i}=\frac{1}{2}$ and $1$, in
section \ref{sec:4} specific heat and magnetic susceptibility data are
presented, in section
\ref{sec:5} the ground state magnetization is considered and finally in section
\ref{sec:6} the conclusions are presented.

\section{Model and Method}
\label{sec:2}
The Hamiltonian for the antiferromagnetic Heisenberg model is
\begin{equation}
H = J \sum_{<i,j>} \textrm{} \vec{s}_{i} \cdot \vec{s}_{j} - h S^{z}
\label{eqn:1}
\end{equation}
where the spins $\vec{s}_{i}$ are located at the vertices of the clusters and
$<i,j>$ denotes nearest neighbors. The coupling constant $J$ is positive and
will be set to $1$ from now on, defining the unit of energy. $h$ is
the strength of an external magnetic field.

Minimization of memory requirements for diagonalization is possible with the
use of the symmetries of the model. These include permutational and spin space
symmetries \cite{Bernu94,Florek02,Waldmann00}. The Hamiltonian commutes with
$S$ and $S^{z}$.
However, even though it is straightforward to work in an $S^{z}$ subspace,
there is no efficient method to construct symmetry adapted eigenstates of
$S$. The Hamiltonian is symmetric under combinations of
permutations of the spins that respect the connectivity of the cluster. The
group of permutations is the symmetry group of the cluster in real space
\cite{Waldmann00}. The model also posseses time-reversal symmetry and inverting
the spins is a symmetry operation in the $S^{z}=0$ sector. The
corresponding group is comprised of the identity and the spin inversion
operation. The full symmetry group of the Hamiltonian is the product of the
space group and the group of spin inversion. Taking the full symmetry
into account, the $S^{z}$ basis states can be projected
into states that transform under specific irreducible representations of the
symmetry group. In this way the Hamiltonian
is block-diagonalized into smaller matrices, and their maximal dimension is
dramatically reduced compared to the full Hilbert space size.
The largest sub-matrix of the block-diagonalized Hamiltonian for the
dodecahedron has dimension $7,058$ for spins $s_{i}=\frac{1}{2}$, therefore
full diagonalization is possible. For $s_{i}=1$ only a few of the lowest
eigenvalues for each irreducible representation were obtained with Lanczos
diagonalization \cite{ARPACK,Shaw}, and for the subspace $S^{z}=1$ this was not
possible for the five-dimensional representations. The largest matrix for which
the lowest eigenvalue was found has dimensionality $13,611,598$ and is
complex.
In the case of the icosahedron full diagonalization is possible for
both $s_{i}=\frac{1}{2}$ and $1$, the largest matrix having a dimension of
$2,982$ in the latter case.

\section{Low Energy Spectra and Correlation Functions}
\label{sec:3}

\subsection{$s_{i}=\frac{1}{2}$}

The low energy spectrum of the icosahedron is presented in table \ref{table:1}.
The ground state energy is a singlet and belongs to the irreducible
representation
$A_{u,s}$, where the first index $g$ or $u$ denotes symmetry or antisymmetry
with respect to space inversion \cite{Altmann94}, and the second $s$ or $a$
symmetry or antisymmetry with respect to spin inversion when $S^{z}=0$. The
energy per spin equals $-0.51566$. The first excited state is also a singlet,
belongs to the representation $H_{g,s}$ and is five-fold degenerate. The next
excited state is also an $S=0$ state, it is non-degenerate and belongs to
$A_{g,s}$. Then a series of $S=1$ states follows. The gap
to the first excited state is $0.53344$ and the singlet-triplet gap $0.89988$.

The nearest neighbor correlation function for the ground state equals
$-0.20626$, almost four times less in strength than the value of an isolated
dimer, $-0.75$. For the first excited state there are two different types of
nearest neighbor correlation functions. The first equals
$-0.16930$ for the pairs $(1,3),(9,11),(5,8),(6,7),(2,10),(4,12)$ in figure
\ref{fig:2}, and the
second $-0.19328$ for the rest of the bonds. The six pairs face each other in
pairs and belong in different triangles. There are five different ways of
distributing the pairs like that on the icosahedron, thus the state is
five-fold degenerate. The second excited singlet has all nearest neighbor
correlation functions equal to $-0.18748$.
The correlation functions other than the nearest neighbor in the ground state
between site $1$ and the rest of the sites are equal to $-0.13908$ for spin
$11$, more than half the value of the nearest neighbor correlation function,
and $0.08408$ for the rest of the spins.


The low energy spectrum of the dodecahedron is shown in table \ref{table:2} and
has the same structure as the one of the icosahedron, after comparison with
table \ref{table:1}. In particular, the four lowest energy
states belong to exactly the same irreducible representations as the
corresponding states in the icosahedron. The spacing of the lowest energy
states is also similar. The ground state energy equals $-9.72219$
\cite{Modine96,NPK01}, and the
energy per spin in the ground state is now higher and equal to $-0.48611$.
Classically, the ground state energy per bond is $-\frac{\sqrt{5}}{3}$ for the
dodecahedron \cite{NPK01} and $-\frac{\sqrt{5}}{5}$ for the icosahedron
(previously reported by Schmidt and Luban in \cite{Schmidt02}),
therefore the energies per spin are equal. Quantum
fluctuations reduce the energy more for the cluster with the highest
coordination number, even though the number of bonds is the same for both. The
gap to the first excited state is also smaller and equal to $0.31567$, and the
same is true for the singlet-triplet gap which is $0.51383$. The nearest
neighbor correlation function is equal to $-0.32407$ for the ground
state \cite{Modine96,NPK01},
quite stronger in magnitude than the one of the icosahedron, as in the
classical case. This is
attributed to the lower coordination number of the dodecahedron. For the first
excited state, the value is $-0.33585$ for the pairs of spins numbered
$(1,2),(9,10),(14,15),(4,12),(17,18),(7,20)$ in figure \ref{fig:1}, while for
the rest of the pairs it is $-0.30797$. The above six pairs are facing each
other in pairs in the dodecahedron and belong in different pentagons. As in the
icosahedron, there are five different ways of distributing the six pairs of
spins in the above manner,
thus the state is five-fold degenerate. The bonds on these pairs are more
singlet like than the ground state, unlike the icosahedron. Another difference
is that the
bonds on the six pairs have now lower energy than the rest of the bonds. For
the next excited state, the nearest neighbor correlation functions are equal to
$-0.31175$.
The correlation functions other than the nearest neighbor in the ground state
between site $1$ and the rest of the sites are equal to $0.06540$ for spins
$3,4,7,8,14$ and $15$, $-0.03882$ for spins $9,10,12,13,16$ and $20$, $0.03307$
for spins $11,17$ and $19$ and $-0.03649$ for spin $18$. They are
significantly smaller than the nearest neighbor correlations. This
is in contrast to the icosahedron, where correlations survive longer distances.


\subsection{$s_{i}=1$}

Full diagonalization is still possible for the icosahedron when $s_{i}=1$.
The low energy spectrum is shown in table \ref{table:3}. The ground state is a
singlet in the $A_{g,s}$ irreducible representation and its energy is
$-18.56111$. It is now symmetric with respect to space inversion, in contrast
to the $s_{i}=\frac{1}{2}$ case. The first excited state is a singlet in the
$A_{u,s}$ representation, which included the ground state for
$s_{i}=\frac{1}{2}$, with energy close to the ground state and equal to
$-18.42539$. The next excited state is a
three-fold degenerate triplet in the representation $T_{2u,a}$ with energy
$-17.83998$, followed by another three-fold degenerate triplet in the
representation $T_{2g,a}$ with energy $-17.80499$. The energy per spin in the
ground state is $-1.54676$ and the nearest neighbor correlation function
$-0.61870$. This value is to be compared with the singlet ground state
energy of a dimer with spins $s_{i}=1$, which equals $-2$. The ratio of the two
values is larger compared with the corresponding ratio for the
$s_{i}=\frac{1}{2}$ case and closer to the classical result. For the first
excited state the nearest neighbor correlation is $-0.61418$, much
closer to the ground
state value compared with the $s_{i}=\frac{1}{2}$ case. The gap to the first
excited state equals $0.13572$ and the singlet-triplet gap $0.72113$.
The correlation functions other than the nearest neighbor in the ground state
between site $1$ and the rest of the sites are equal to $-0.74630$ for spin
$11$, and $0.36796$ for the rest of the spins. Compared with the
$s_{i}=\frac{1}{2}$ case, the magnitudes of the next than nearest neighbor
correlation functions are significantly increased with respect to nearest neighbors,
and the correlation with
site $11$ is even stronger than the nearest neighbor correlation.


As was the case for $s_{i}=\frac{1}{2}$, the low energy spectrum (table
\ref{table:4}) is similar for the clusters when $s_{i}=1$. The ground state
for the dodecahedron is a non-degenerate singlet with energy $-30.24551$, and
an energy per spin equal to $-1.51228$. Again the energy per spin is higher
than the one of the icosahedron. The first excited state is also a singlet
close to the ground state with energy $-30.21750$, and the next two excited
states are triply degenerate triplets with very close energies, $-29.92161$ and
$-29.91011$. The energy gap is $0.02801$ and the singlet-triplet gap $0.32390$,
values again smaller than the ones of the icosahedron.
The nearest neighbor correlation functions in the ground and first excited
states are equal to $-1.00818$ and $-1.00725$ respectively.
The correlation functions other than the nearest neighbor in the ground state
between site $1$ and the rest of the sites are equal to $0.27896$ for spins
$3,4,7,8,14$ and $15$, $-0.20912$ for spins $9,10,12,13,16$ and $20$, $0.35962$
for spins $11,17$ and $19$ and $-0.47333$ for spin $18$. Similarly to the
icosahedron, the magnitudes are increased with respect to the nearest neighbor
correlations compared with the $s_{i}=\frac{1}{2}$ case. Farther than nearest
neighbor correlations are stronger for the icosahedron, as was the case for
$s_{i}=\frac{1}{2}$. It wasn't possible to diagonalize the five-fold degenerate
irreducible representations for $S^{z}=1$. Therefore the $S$ values in
parentheses in table \ref{table:4} are deduced by comparison with the states of
the icosahedron, since the spectra are similar. In any case these $S$ values
can only be $0$ or $1$, with these energies absent from the $S^{z}=2$
spectrum.

\section{Specific Heat and Magnetic Susceptibility}
\label{sec:4}
\begin{figure}
\includegraphics[width=5in,height=3.1in]{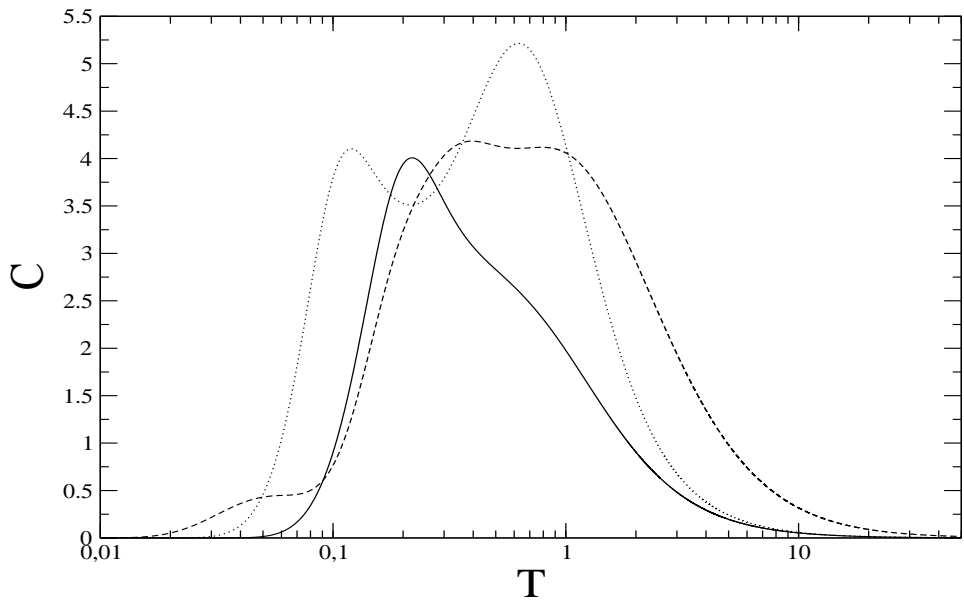}
\caption{Specific heat $C$ of the antiferromagnetic Heisenberg model as a
function of temperature $T$. Solid line: icosahedron with $s_{i}=\frac{1}{2}$,
dotted line: dodecahedron with $s_{i}=\frac{1}{2}$, dashed line: icosahedron
with $s_{i}=1$. $C$ is in arbitrary units and $T$ in units of energy.}
\label{fig:3}
\end{figure}

\begin{figure}
\includegraphics[width=5in,height=3.1in]{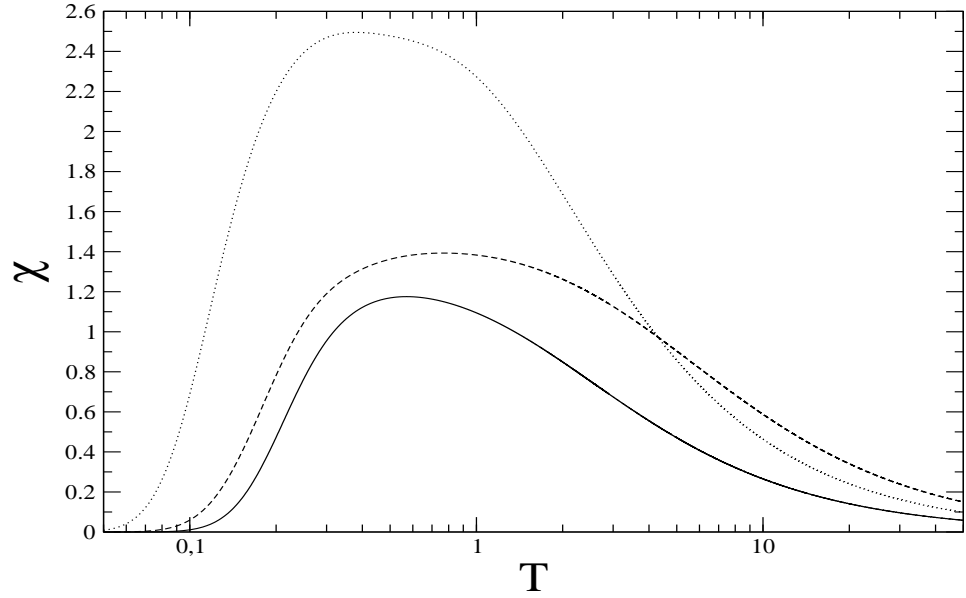}
\caption{Magnetic susceptibility $\chi$ of the antiferromagnetic Heisenberg
model as a function of temperature $T$. Solid line: icosahedron with
$s_{i}=\frac{1}{2}$, dotted line: dodecahedron with $s_{i}=\frac{1}{2}$, dashed
line: icosahedron with $s_{i}=1$. $\chi$ is in arbitrary units and $T$ in units
of energy.}
\label{fig:4}
\end{figure}

The temperature dependence of the specific heat and the magnetic susceptibility
for the cases where
exact diagonalization is possible is shown in figures \ref{fig:3} and
\ref{fig:4} respectively.
For the icosahedron and $s_{i}=\frac{1}{2}$, there is a peak in the specific
heat around $T=0.219$ and a shoulder around $T=0.8$. The
peak is inside the energy gap, a feature characteristic of frustrated systems
\cite{Sindzingre00,Syromyatnikov03}.
For the dodecahedron there are two
well defined peaks. The first peak is centered around $T=0.120$ and the second
around $T=0.627$. The first peak is inside the energy gap. Similar results have
been obtained for other frustrated systems, and Sindzingre {\it et al.}
attributed the peak to the combined effect of singlet excitations inside the
singlet-triplet gap and low-lying triplet excitations for the Kagom\'e
lattice \cite{Sindzingre00}. In contrast, Syromyatnikov and Maleyev
considered a Kagom\'e star where the peak results from the increase in
the density of states just above the spin gap
\cite{Syromyatnikov03}. In the present case both the $S=0$ and $S=1$
sectors contribute for the peak inside the energy gap. The $S=0$
contribution is due to the low-lying singlets, while the $S=1$
contribution is also due to a few of the lowest energy triplets. For
the shoulder and the higher energy peak in the two systems
respectively higher $S$ sectors contribute as well.
For the icosahedron and $s_{i}=1$, there are three regions in
the graph. At temperatures lower than the energy gap, the specific heat
initially rises slowly towards an area between $T \approx 0.055$ to $0.07$
where it almost stabilizes with temperature. The contributions come
from the lowest energy states in the $S=0$ and $S=1$ sectors. The
specific heat then increases rapidly
reaching a peak around $T=0.395$, followed by a slight decrease to a local
minimum at a temperature around the singlet-triplet gap,
$T=0.619$. Following that it increases again to another peak around
$T=0.786$ and then starts dropping towards zero. The contributions to the
two peaks and the local minimum come from various spin sectors. Even
though the double-peak structure is similar to the specific heat of
the dodecahedron for $s_{i}=\frac{1}{2}$, there is no peak inside the
energy gap, and this signifies a change with respect to the
$s_{i}=\frac{1}{2}$ case. The magnetic susceptibility is plotted in figure
\ref{fig:4}, with a temperature dependence similar for the three
cases. The position of the maximum follows the pattern for the lowest
temperature peak of the specific heat, with a peak temperature
decreasing with size and increasing with spin magnitude.

\section{Ground State Magnetization}
\label{sec:5}
\begin{figure}
\includegraphics[width=5in,height=3.5in]{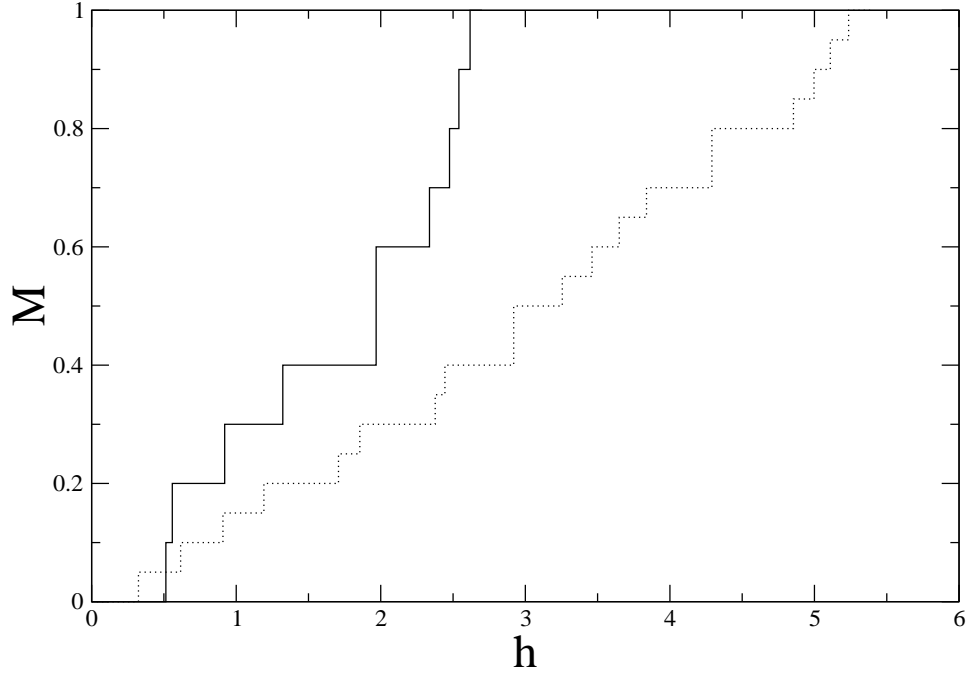}
\caption{Ground state magnetization $M$ as a function of magnetic field $h$ for
the dodecahedron. $M$ is the total spin $S$ normalized to the number of sites
and the magnitude of spin. Solid line: $s_{i}=\frac{1}{2}$, dotted line:
$s_{i}=1$. $M$ has no units and $h$ is in units of energy. The steps between
$S$ sectors are $0.1$ for $s_{i}=\frac{1}{2}$ and $0.05$ for $s_{i}=1$. The
discontinuities are between $0.4$ and $0.6$ for $s_{i}=\frac{1}{2}$, and
between $0.4$ and $0.5$, and $0.7$ and $0.8$ for $s_{i}=1$.}
\label{fig:5}
\end{figure}

Frustrated spin systems have been found to exhibit magnetization jumps in the
presence of an external field, where the lowest state in a particular $S$
sector never becomes the ground state in the field \cite{Richter04}.
The magnetization in an external field has been calculated for the dodecahedron
at the classical level, and such a discontinuity was found \cite{Coffey92}. A
discontinuity in the magnetization was also found when the classical state is
perturbed with the quantum fluctuations \cite{NPK01}. The calculation for the
classical ground state of the icosahedron shows also the presence of a
magnetization jump in a field. In the quantum case, the lowest energy
in a magnetic field in a particular $S$ sector is calculated from the
lowest energy in the absence of a field by adding the Zeeman term. The lowest
energies in the different $S$ sectors are shown in tables
\ref{table:5} and \ref{table:6} for the dodecahedron for
$s_{i}=\frac{1}{2}$ and $s_{i}=1$ respectively.
As was mentioned before, calculation of the lowest energies is not
possible for the five-dimensional irreducible representations when $S^{z}=1$.
However, by comparing the spectra of the five-dimensional representations for
$S^{z}=0$ with the spectra of the rest of the representations for $S^{z}=1$, it
is seen that the lowest energy for $S=1$ does not belong to the
five-dimensional representations (table \ref{table:4}). The ground state
magnetization is found by comparing the energies in a field in the different
sectors. The plots for the dodecahedron for both
$s_{i}=\frac{1}{2}$ and $1$ are shown in figure \ref{fig:5}, where the
magnetization $M$ is the total spin $S$ normalized to the number of
sites and the magnitude of each
spin, with steps between different $S$ sectors equal to $0.1$ for
$s_{i}=\frac{1}{2}$ and $0.05$ for $s_{i}=1$. There is a
discontinuity of $M$ between $0.4$ and $0.6$ for
$s_{i}=\frac{1}{2}$ ($S=4$ and $6$), and two discontinuities for
$s_{i}=1$, between $0.4$ and
$0.5$ ($S=8$ and $10$), and between $0.7$ and $0.8$ ($S=14$ and
$16$). These discontinuities are associated with magnetization
plateaux on both sides of the jumps \cite{Honecker04}. Magnetization plateaux
are also observed in figure \ref{fig:5} for $M=0$ when
$s_{i}=\frac{1}{2}$, and for $M=0.2$ and $0.3$ for $s_{i}=1$.

As seen from tables \ref{table:5} and \ref{table:6}, the ground states
on each side of the jumps are non-degenerate, and the spatial symmetry
of the ground state switches from symmetric to antisymmetric as the
field increases. Spin inversion symmetry has only be determined for
$s_{i}=\frac{1}{2}$ for the cases of interest, and in that case this
symmetry does not change as the discontinuity takes place. The lowest
state of $M$ that never becomes the ground state is in every case
symmetric with respect to spatial inversion, and it belongs to the same
irreducible representation for the $s_{i}=\frac{1}{2}$ and the lowest
field $s_{i}=1$ jump. It is also interesting that in the $s_{i}=1$
case, even though the lowest
state for $S=6$ is non-degenerate as is the $S=8$ state, there is no
discontinuity associated with these two states, indicating the
significance of the different spatial inversion symmetry for the
states on the two sides of a magnetization jump. However, a
magnetization plateau exists for the corresponding magnetization
$M=0.3$.

The behavior of the correlation functions on either side of the
magnetization jumps is shown in table \ref{table:7}. The
nearest-neighbor and next nearest-neighbor correlation functions, as
well as $\vec{S}_{1} \cdot \vec{S}_{11}$, become more positive as the
magnetic field is getting stronger. For $\vec{S}_{1} \cdot \vec{S}_{18}$, the
magnetization jump decreases its value with increasing field. However,
this weakening is very pronounced for the $S=8$ to $S=10$
transition for $s_{i}=1$.
For $\vec{S}_{1} \cdot \vec{S}_{9}$, the lower jump in the
$s_{i}=1$ case decreases its strength, contrary to what happens for
the other two discontinuities. It is concluded that the first
discontinuity in the $s_{i}=1$ case has different characteristics from
the other two jumps as far as longer range correlation functions are
concerned. This could possibly be due to the different change in
behavior with respect to spin inversion as this discontinuity takes
place, compared with the other two discontinuities.

The data for the two lowest quantum numbers $s_{i}$ shows that the
discontinuity in the classical solution survives the quantum fluctuations. The
presence of jumps for $s_{i}=\frac{1}{2},1$ and $\infty$
indicates that this is probably a characteristic of the
system for any quantum number. The common features of the mechanism of
the effect for the two lowest quantum numbers also point in this
direction. For the icosahedron there is no such jump for
$s_{i}=\frac{1}{2}$ or $1$, even though there is a jump at the classical level,
and quantum fluctuations are stronger.
It is therefore concluded that the discontinuity in the magnetization at the
quantum level is not a consequence of the symmetry, but rather of the polygon
structure of the clusters. For larger clusters with non-equivalent sites and
including hexagons no such discontinuity was found \cite{NPK05-2}. Thus the
pentagon-only structure of the dodecahedron appears crucial for the presence of
the discontinuities.


\section{Conclusions}
\label{sec:6}
The antiferromagnetic Heisenberg model was solved on the icosahedron and
the dodecahedron, which belong to the point symmetry group $I_{h}$. It was
found that the low energy spectra are similar, and consist of excited states
that are singlets for both $s_{i}=\frac{1}{2}$ and $1$. Spin correlations
survive longer distances for the icosahedron. The specific heat calculation
revealed a non-trivial dependence on temperature, with more than one peak
for the dodecahedron for $s_{i}=\frac{1}{2}$, and the icosahedron for
$s_{i}=1$.
This feature has also been found for other
frustrated spin systems \cite{Sindzingre00}-\cite{Kubo93}. There are
discontinuities in the ground state magnetization as a function of magnetic
field
for the dodecahedron for both $s_{i}=\frac{1}{2}$ and $1$, but this is not the
case for the icosahedron. This discontinuity was found at the classical level
\cite{Coffey92}, and here it has been shown to survive the quantum
fluctuations. It appears to be a characteristic of the system for any quantum
number. There is also a number of magnetization plateaux for both quantum
numbers. These non-trivial properties of the clusters are a consequence of
the frustrated interactions and the connectivity. Other clusters of
fullerene-type geometry show similar properties in their low energy spectrum
\cite{NPK05-2}. The similarity in the spectra and the specific heat of the
icosahedron and the dodecahedron points in similar properties for the larger
cluster with the same symmetry and $60$ sites, also equivalent. It also
shows that more complicated fermionic models for the $60$-site system,
such as the Hubbard model, could instead be considered on the
icosahedron and the dodecahedron, and the results will be reliable predictions
for the properties of the models on the larger system. However, no
prediction can be made for the response in a field for the antiferromagnetic
Heisenberg model, which depends not only on the symmetry but also on the
polygons that make up the clusters. The dodecahedron exhibits a jump and has
only pentagons, while the $60$-site system has hexagons as well. It has been
found that hexagon correlations are more important for frustrated systems of
the fullerene type with more than $20$ sites, for which there is no
magnetization discontinuity \cite{NPK05-2}. Another point deserving further
attention is the transition from the quantum to the classical limit,
regarding the low energy spectra and the presence of singlets, the
temperature dependence of the specific heat and the presence of
magnetization discontinuitites.
The general behavior of the clusters with
$I_{h}$ point group symmetry is determined to a significant extent by the
symmetry, however the coordination number and the polygons that make up the
structures are also important for the determination of their properties.



The author thanks D. Coffey and P. Herzig for discussions. Calculations were
carried out at the Trinity Center for High Performance Computing. The work was
supported by a Marie Curie Fellowship of the European Community program
Development Host Fellowship under contract number HPMD-CT-2000-00048.

$^{*}$ Present address: Leoforos Syggroy 360, Kallithea 176 74,
Athens, Hellas.

\bibliography{paperfour}

\newpage

%
%


\begin{table}[h]
\begin{center}
\caption{Low energy spectrum for the icosahedron for $s_{i}=\frac{1}{2}$. The
first index of the irreducible representations indicates the behavior under
spatial inversion, where $g$ stands for symmetric and $u$ for antisymmetric
\cite{Altmann94}, and the second under spin inversion for the $S^{z}=0$
component, where $s$ stands for symmetric and $a$ for antisymmetric.}
\begin{tabular}{c|c|c|c|c|c|c|c}
energy & multiplicity & irreducible & $S$ & energy & multiplicity & irreducible
& $S$ \\
& & representation & & & & representation &\\
\hline
-6.18789 & 1 & $A_{u,s} $ & 0 & -4.82887 & 5 & $H_{u,s}$  & 0 \\
\hline
-5.65445 & 5 & $H_{g,s} $ & 0 & -4.76398 & 5 & $H_{u,a}$  & 1 \\
\hline
-5.62426 & 1 & $A_{g,s} $ & 0 & -4.50000 & 4 & $F_{g,s}$  & 0 \\
\hline
-5.28801 & 3 & $T_{2g,a}$ & 1 & -4.50000 & 1 & $A_{g,s}$  & 0 \\
\hline
-5.17728 & 3 & $T_{1u,a}$ & 1 & -4.44972 & 4 & $F_{u,a}$  & 1 \\
\hline
-5.10989 & 3 & $T_{2u,a}$ & 1 & -4.31691 & 5 & $H_{g,a}$  & 1 \\
\hline
-4.86352 & 4 & $F_{g,a}$  & 1 & -4.29999 & 3 & $T_{2u,a}$ & 1 \\
\end{tabular}
\label{table:1}
\end{center}
\end{table}


\begin{table}[h]
\begin{center}
\caption{Low energy spectrum for the dodecahedron for $s_{i}=\frac{1}{2}$.
Notation as in table \ref{table:1}.}
\begin{tabular}{c|c|c|c|c|c|c|c}
energy & multiplicity & irreducible & $S$ & energy & multiplicity & irreducible
& $S$ \\
& & representation & & & & representation & \\
\hline
-9.72219 & 1 & $A_{u,s} $ & 0 & -8.87964 & 5 & $H_{u,s}$  & 0 \\
\hline
-9.40651 & 5 & $H_{g,s} $ & 0 & -8.69499 & 3 & $T_{2u,a}$ & 1 \\
\hline
-9.35236 & 1 & $A_{g,s} $ & 0 & -8.69441 & 5 & $H_{u,a}$  & 1 \\
\hline
-9.20836 & 3 & $T_{2g,a}$ & 1 & -8.66571 & 3 & $T_{1u,a}$ & 1 \\
\hline
-9.18649 & 4 & $F_{u,a}$  & 1 & -8.65030 & 4 & $F_{g,s}$  & 2 \\
\hline
-9.13048 & 3 & $T_{2u,a}$ & 1 & -8.63614 & 5 & $H_{g,a}$  & 1 \\
\hline
-8.97112 & 3 & $T_{1g,a}$ & 1 & -8.63178 & 4 & $F_{g,a}$  & 1 \\
\end{tabular}
\label{table:2}
\end{center}
\end{table}


\begin{table}[h]
\begin{center}
\caption{Low energy spectrum for the icosahedron for $s_{i}=1$. Notation as in
table \ref{table:1}.}
\begin{tabular}{c|c|c|c|c|c|c|c}
energy & multiplicity & irreducible & $S$ & energy & multiplicity & irreducible
& $S$\\
& & representation & & & & representation & \\
\hline
-18.56111 & 1 & $A_{g,s}$  & 0 & -17.15212 & 3 & $T_{1g,a}$ & 1 \\
\hline
-18.42539 & 1 & $A_{u,s}$  & 0 & -17.00416 & 5 & $H_{g,s}$  & 0 \\
\hline
-17.83998 & 3 & $T_{2u,a}$ & 1 & -16.97485 & 4 & $F_{u,a}$  & 1 \\
\hline
-17.80499 & 3 & $T_{2g,a}$ & 1 & -16.82345 & 5 & $H_{g,a}$  & 1 \\
\hline
-17.60137 & 5 & $H_{u,s}$  & 0 & -16.75333 & 5 & $H_{u,a}$  & 1 \\
\hline
-17.19717 & 4 & $F_{g,s}$  & 0 & -16.74705 & 1 & $A_{g,s}$  & 0 \\
\hline
-17.17453 & 5 & $H_{g,s}$  & 0 & -16.73750 & 4 & $F_{g,a}$  & 1 \\
\end{tabular}
\label{table:3}
\end{center}
\end{table}


\begin{table}[h]
\begin{center}
\caption{Low energy spectrum for the dodecahedron for $s_{i}=1$. Notation as in
table \ref{table:1}. The $S$ numbers in parentheses can be $0$ or $1$ since
they are missing from the $S^{z}=2$ spectrum, and they are assigned values
after comparison with the icosahedron spectrum in table \ref{table:3}.}
\begin{tabular}{c|c|c|c|c|c|c|c}
energy & multiplicity & irreducible & $S$ & energy & multiplicity & irreducible
& $S$ \\
& & representation & & & & representation & \\
\hline
-30.24551 & 1 & $A_{g,s}$  &   0 & -29.61145 & 3 & $T_{1g,a}$ &   1 \\
\hline
-30.21750 & 1 & $A_{u,s}$  &   0 & -29.57332 & 4 & $F_{g,s}$  &   0 \\
\hline
-29.92161 & 3 & $T_{2u,a}$ &   1 & -29.50512 & 3 & $T_{1u,a}$ &   1 \\
\hline
-29.91011 & 3 & $T_{2g,a}$ &   1 & -29.45063 & 4 & $F_{g,a}$  &   1 \\
\hline
-29.85881 & 5 & $H_{g,s}$  & (0) & -29.39457 & 5 & $H_{u,a}$  & (1) \\
\hline
-29.67223 & 5 & $H_{u,s}$  & (0) & -29.35464 & 5 & $H_{g,a}$  & (1) \\
\hline
-29.65951 & 4 & $F_{u,a}$  &   1 & -29.31754 & 3 & $T_{2u,a}$ &   1 \\
\end{tabular}
\label{table:4}
\end{center}
\end{table}

\begin{table}[h]
\begin{center}
\caption{Lowest energy in the various $S$ sectors for the dodecahedron for
$s_{i}=\frac{1}{2}$. Notation as in table \ref{table:1}.}
\begin{tabular}{c|c|c|c|c|c|c|c}
$S$ & energy & multiplicity & irreducible & $S$ & energy & multiplicity &
irreducible \\
& & & representation & & & & representation \\
\hline
 0 & -9.72219 & 1 & $A_{u,s}$  &  6 & -2.47099 & 1 & $A_{u,s}$  \\
\hline
 1 & -9.20836 & 3 & $T_{2g,a}$ &  7 & -0.13397 & 3 & $T_{1u,a}$ \\
\hline
 2 & -8.65030 & 4 & $F_{g,s}$  &  8 &  2.34152 & 5 & $H_{g,s}$  \\
\hline
 3 & -7.72967 & 4 & $F_{g,a}$  &  9 &  4.88197 & 3 & $T_{2u,a}$ \\
\hline
 4 & -6.40730 & 1 & $A_{g,s}$  & 10 &  7.5     & 1 & $A_{g,s}$  \\
\hline
 5 & -4.37206 & 4 & $F_{g,a}$ \\
\end{tabular}
\label{table:5}
\end{center}
\end{table}

\begin{table}[h]
\begin{center}
\caption{Lowest energy in the various $S$ sectors for the dodecahedron for
$s_{i}=1$. Notation as in table \ref{table:1}. For sectors with $S$ higher than
1 the symmetry under spin inversion for the $S^{z}=0$ component has not been
determined (except for $S=20$).}
\begin{tabular}{c|c|c|c|c|c|c|c}
$S$ & energy & multiplicity & irreducible & $S$ & energy & multiplicity &
irreducible \\
& & & representation & & & & representation \\
\hline
 0 & -30.24551 & 1 & $A_{g,s}$  & 11 & -9.72811 & 3 & $T_{2g}$  \\
\hline
 1 & -29.92161 & 3 & $T_{2u,a}$ & 12 & -6.26689 & 5 & $H_{u}$   \\
\hline
 2 & -29.30598 & 5 & $H_{g}$    & 13 & -2.61717 & 4 & $F_{g}$   \\
\hline
 3 & -28.39788 & 3 & $T_{1u}$   & 14 &  1.22128 & 1 & $A_{g}$   \\
\hline
 4 & -27.20612 & 4 & $F_{g}$    & 15 &  5.52324 & 3 & $T_{2g}$  \\
\hline
 5 & -25.49868 & 4 & $F_{g}$    & 16 &  9.80295 & 1 & $A_{u}$   \\
\hline
 6 & -23.64370 & 1 & $A_{g}$    & 17 & 14.65742 & 3 & $T_{1u}$  \\
\hline
 7 & -21.26717 & 4 & $F_{g}$    & 18 & 19.65432 & 5 & $H_{g}$   \\
\hline
 8 & -18.82343 & 1 & $A_{g}$    & 19 & 24.76393 & 3 & $T_{2u}$  \\
\hline
 9 & -15.88505 & 4 & $F_{g}$    & 20 & 30       & 1 & $A_{g,s}$ \\
\hline
10 & -12.98367 & 1 & $A_{u}$ \\
\end{tabular}
\label{table:6}
\end{center}
\end{table}

\begin{table}[h]
\begin{center}
\caption{Unique ground state correlation functions for the lowest energy states
on either side of the magnetization discontinuities.}
\begin{tabular}{c|c|c|c|c|c|c}
correlation & $s_{i}=\frac{1}{2}$ & $s_{i}=\frac{1}{2}$ & $s_{i}=1$ & $s_{i}=1$
& $s_{i}=1$ & $s_{i}=1$ \\
function & $S=4$ & $S=6$ & $S=8$ & $S=10$ & $S=14$ & $S=16$ \\
\hline
$\vec{S}_{1} \cdot \vec{S}_{2}$ & -0.21358 & -0.08237 & -0.62745 & -0.43279 &
 0.04071 & 0.32677 \\
\hline
$\vec{S}_{1} \cdot \vec{S}_{3}$ & 0.08764 & 0.12061  & 0.34296 & 0.43033 &
0.60512 & 0.72748 \\
\hline
$\vec{S}_{1} \cdot \vec{S}_{9}$ & 0.04971 & 0.08038 & 0.18218 & 0.16926 &
0.47185 & 0.59169 \\
\hline
$\vec{S}_{1} \cdot \vec{S}_{11}$ & -0.01650 & 0.12180 & -0.04887 & 0.36408 &
0.45061 & 0.73273 \\
\hline
$\vec{S}_{1} \cdot \vec{S}_{18}$ & 0.11611 & 0.02574 & 0.47814 & 0.10860 &
0.56423 & 0.50655 \\
\end{tabular}
\label{table:7}
\end{center}
\end{table}

\newpage

\end{document}